\documentclass{article}
\usepackage{ijcai24}

\usepackage{times}
\usepackage{soul}
\usepackage{url}
\usepackage[utf8]{inputenc}
\usepackage[small]{caption}
\usepackage{graphicx}
\usepackage{amsmath}
\usepackage{amsthm}
\usepackage{booktabs}
\usepackage{algorithm}
\usepackage{algorithmic}
\usepackage[switch]{lineno}

\usepackage[hidelinks]{hyperref}

\title{False Sense of Security in Explainable Artificial Intelligence (XAI)}

\author{
Neo Christopher Chung$^{1,*}$\and
Hongkyou Chung$^{2,3}$\and
Hearim Lee$^{4}$\and\\
Lennart Brocki$^{1}$\and
Hongbeom Chung$^{2,5}$\And
George Dyer$^{6,*}$
\affiliations
$^1$Institute of Informatics, University of Warsaw\\
$^2$School of Law, Seoul National University, Seoul, Republic of Korea\\
$^3$Shin $\&$ Kim LLC, Seoul, Republic of Korea\\
$^4$Dongin Law Group, Seoul, Republic of Korea\\
$^5$Wesang Co., Ltd., Seoul, Republic of Korea\\
$^6$Informatism LLC, New Mexico, United States\\
\emails
$^{*}$n.chung@uw.edu.pl, george@informatism.com
}

\begin{document}

\maketitle

\begin{abstract}
    A cautious interpretation of AI regulations and policy in the EU and the USA places explainability as a central deliverable of compliant AI systems. However, from a technical perspective, explainable AI (XAI) remains an elusive and complex target where even state-of-the-art methods often reach erroneous, misleading, and incomplete explanations. ``Explainability'' has multiple meanings which are often used interchangeably, and there are an even greater number of XAI methods – none of which presents a clear edge. Indeed, there are multiple failure modes for each XAI method, which require application-specific development and continuous evaluation. In this paper, we analyze legislative and policy developments in the United States and the European Union, such as the Executive Order on the Safe, Secure, and Trustworthy Development and Use of Artificial Intelligence, the AI Act, the AI Liability Directive, and the General Data Protection Regulation (GDPR) from a right to explanation perspective. We argue that these AI regulations and current market conditions threaten effective AI governance and safety because the objective of trustworthy, accountable, and transparent AI is intrinsically linked to the questionable ability of AI operators to provide meaningful explanations. Unless governments explicitly tackle the issue of explainability through clear legislative and policy statements that take into account technical realities, AI governance risks becoming a vacuous ``box-ticking'' exercise where scientific standards are replaced with legalistic thresholds, providing only a false sense of security in XAI.
\end{abstract}

\begin{figure*}[tb!]
\centering
  \includegraphics[width=.9\textwidth]{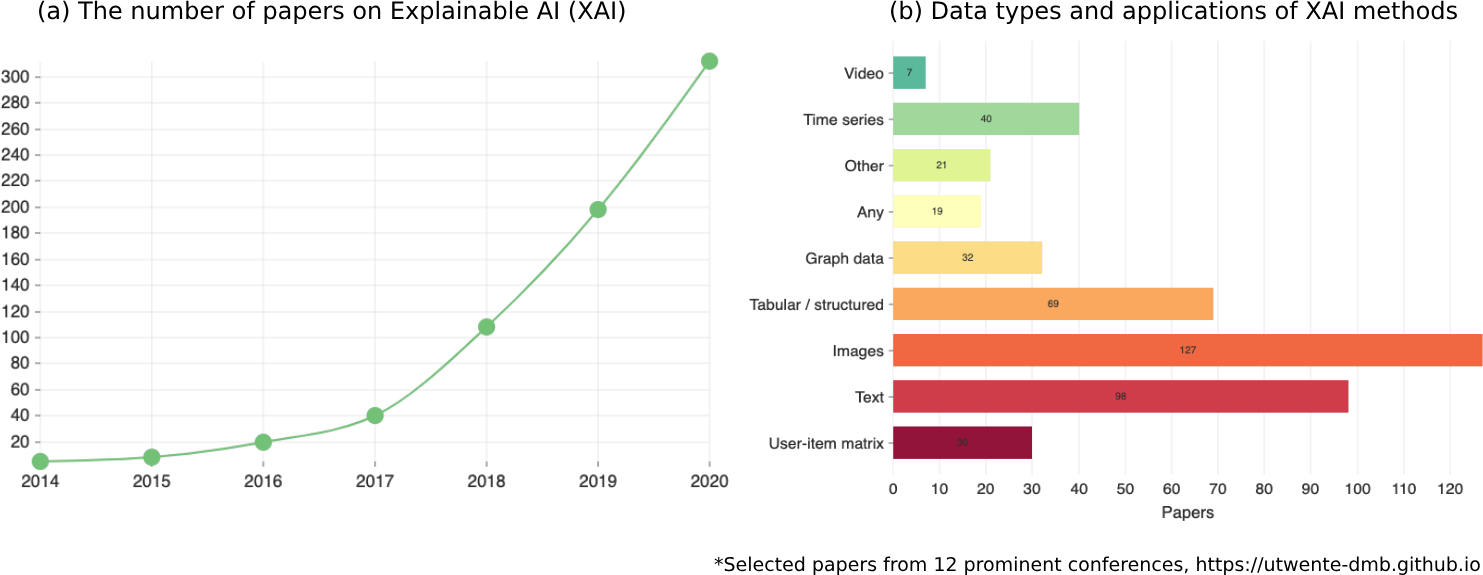}
  \caption{Explosive growth of XAI research. The number of XAI papers in 12 AI/ML conferences are counted and categorized by Nauta \emph{et al.}, 2023.}\label{fig:xai-papers}
\end{figure*}

\section{Introduction}

Recent advancements in AI rely on end-to-end training of deep neural networks (DNNs), which act like black box algorithms. While the superb performance of predictive and generative models led to their wide applications in all areas of society, from social media to healthcare, developers, and users do not have a full grasp of their operating characteristics. The output (e.g., classification of an image) is obtained without understanding the AI's decision-making process. Yet, governance of AI demands revealing and reporting this mechanism as much as possible. Explanations are necessary because fundamentally, humans need to understand automated decision-making for meaningful governance. Furthermore, laws are being proposed and enacted to demand that AI companies provide such explanations.
 
There’s a general consensus that increasing explainability will lead to transparent, reliable, and accountable AI, which contributes to greater safety. Yet, these terms -- such as explainability, interpretability, and transparency -- remain polysemantic and under-defined. These nuances are lost in recent legislative and policy texts leaving a large gap between technical feasibility and practical implementation on the one hand, and compliance thresholds on the other. Furthermore, these upcoming laws have prompted for-profit companies to eagerly market various XAI, governance, and risk management solutions that are half-baked. In the current landscape, we are concerned that the explosive development of XAI could paradoxically lead to a false sense of security. 

There has been a rapid growth in research and development in XAI \cite{Jacovi2023}. What started as a niche topic a decade ago has become prominent in all major machine learning (ML) and AI conferences, including specialized workshops on XAI (IJCAI, ICML, ECAI, and others). A dedicated conference on XAI, called the World Conference on Explainable Artificial Intelligence, started last year \cite{xAIconf2023}. Overall, the number of papers studying and developing XAI has increased exponentially within the top 12 ML and AI conferences \cite{Nauta2023}. Including a wide range of journals, workshops, and other venues would increase this number drastically. While there are attempts to categorize and taxonomize assumptions and aims of XAI under development, often a bird eye view on different paradigms of explainability is lacking in the field of AI governance and policy.

\section{Legislative and Policy Developments}

 Generally, AI policy statements focus on making AI safe, secure, transparent, trustworthy, and explainable. These concepts, however, receive less attention or detailed specification in the laws enacted to implement these policies. This led to ongoing debate about the extent to which a right to explanation exists. Based on relevant developments, we argue that some form of explainability will likely become a \emph{de facto} requirement for critical AI systems deployed on the global market.

\subsection{United States of America}

In the United States, Joe Biden's October 2023 Executive Order on the Safe, Secure, and Trustworthy Development and Use of Artificial Intelligence (14110) calls upon the executive agencies under his authority to study and propose regulations that can effectively govern the safe and transparent use of AI in both the private and public sectors \cite{EO}. The National Institute of Standards and Technology (NIST), Department of Energy (DoE), National Science Foundation (NSF), and other agencies are crafting relevant resources, guidelines, and plans with industry and stakeholders.

Yet, the executive order is largely a statement of policy as it has no binding effect on deployers of AI systems in the U.S. The order mentions ``safe'', ``secure'', and their derivative words around 80 times. ``Trust'' and its derivatives appear a dozen times while ``transparent'' and associated terms are mentioned four times. Explainability is referred to only once, in the context of the U.S. government's need for ``emphasizing or clarifying requirements and expectations related to the transparency of AI models and regulated entities’ ability to explain their use of AI models.''

Under U.S. law, there is currently only one binding directive relating to explainability: the Equal Credit Opportunity Act (ECOA), which requires creditors to notify applicants who are denied credit with specific reasons for an adverse credit decision. Specifically, §1002.9(b)(2) provides:

\begin{quote}
The statement of reasons for adverse action required by paragraph (a)(2)(i) of this section must be specific and indicate the principal reason(s) for the adverse action. Statements that the adverse action was based on the creditor's internal standards or policies or that the applicant, joint applicant, or similar party failed to achieve a qualifying score on the creditor's credit scoring system are insufficient\cite{EqualCredit15/1691}.
\end{quote}

Enacted in 1974, the ECOA was not written with generative AI in mind, but rather the deterministic decision-making systems in use at the time. Applied to contemporary technology, it nonetheless sets an actionable standard by ruling out the use of any black box AI systems that do not permit the isolation of the primary factors that contributed to their output. Such a clear-line rule is effective and practicable, but only because the law applies to a narrow scope of activity. It remains to be seen whether the continuation of this piecemeal legislative approach to the use of algorithmic systems is desirable, as the multiplication of possible use cases of AI will almost certainly outpace the ability of the U.S. legislature to address them. At some point, broader regulations will be needed.

\subsection{European Union}

Broader and more ambitious initiatives are now being promulgated within Europe: the AI Act \cite{AIAct}, the General Data Protection Regulation (GDPR) \cite{GDPR}, and the AI Liability Directive \cite{AILiability}. Like the GDPR, the AI Act will be a regulation when it comes into force over the coming months. This means that they are considered as controlling law in every country within the EU. A directive, on the other hand, directs member states to adopt legislation that follows its principles, which may result in some differences among the implementing laws of each member state \cite{EULawTypes}.

\subsubsection{AI Act}

The AI Act was endorsed by members of the European Parliament (MEP) in March 2024 \cite{AIActMEP}, marking the first foray into the generalized regulation of AI. The Act and its relative merits have been discussed in detail elsewhere, and for the purposes of this paper, it is sufficient to note that it does not promulgate any substantive and precise explainability thresholds for AI developers. In fact, the AI Act itself does explicitly not reference explainability, focusing instead on the need for transparency in high-risk AI models.

Transparency is addressed in Article 13(1) \cite{AIAct}, which states that ``High-risk AI systems shall be designed and developed in such a way to ensure that their operation is sufficiently transparent to enable users to interpret the system’s output and use it appropriately. An appropriate type and degree of transparency shall be ensured.'' The ethics guidelines of the European Commission (2019) clarify this requirement insofar as the decision-making process must be recorded and documented, and an ``understandable explanation
of the algorithmic decision-making processes'' must be published \cite{AIEthics}. According to Articles 6 and 7, high-risk systems would include those deployed within the education, health, recruitment, critical infrastructure management, law enforcement, and justice domains because such use cases could result in threats to the health, safety, or fundamental rights of individuals \cite{AIAct}.

As written, the AI Act does not provide any tangible thresholds for the transparency of high-risk models. Instead, the specific technical requirements of the models affected by Article 13 will be defined through a collaborative approach by the governing institutions created by the Act: the AI Office, the European Artificial Intelligence Board, the Advisory Forum, and the Scientific Panel of Independent Experts. This decentralized approach is commendable because it represents a tacit admission from the European legislators that the issue of AI regulation is too complex for a single body, much less a political one, to take on alone. Only through a deliberative (and almost certainly iterative) process will specific thresholds be defined for high-risk AI models. 

On the other hand, the silence on the issue of explainability is deafening. After several years of study and deliberation, the inability to seriously address the concept does not bode well for future efforts, possibly camouflaging its intractable issues. More precise yet possibly misaligned guidance from the AI Act may have been more useful than no guidance at all, as stakeholders across the globe are having to guess as to how their models may or may not meet future EU regulatory standards. One particular area of uncertainty is whether an expectation of explainability will not apply to non-high-risk models in Europe. Such an assumption could be misguided, because a cautious reading of the GDPR and the AI Liability Directive shows that explainability will be – in some shape or form – a \emph{de facto} compliance requirement for all AI models deployed in the EU. 

\subsubsection{General Data Protection Regulation (GDPR)}

To ensure fair and transparent processing of personal data, GDPR Articles 13, 14, and 15 require the controller to provide the prospective or existing data subject with information relating to: ``the existence of automated decision-making, including profiling, referred to in Article 22(1) and (4) and, at least in those cases, meaningful information about the logic involved, as well as the significance and the envisaged consequences of such processing for the data subject'' \cite{GDPR}. Processing is defined in Article 4(2) as ``any operation or set of operations which is performed on personal data or on sets of personal data, whether or not by automated means.''

A narrow right to explanation potentially exists within GDPR Article 22 (3) insofar as it requires the data controller to ``implement suitable measures to safeguard the data subject's rights and freedoms and legitimate interests, at least the right to obtain human intervention on the part of the controller, to express his or her point of view and to contest the decision,'' when an individual is submitted to automated processing through contract or consent. Since Articles 13, 14, and 15 give a person a right to knowledge about the ``logic involved'' in the processing they are subject to, it follows that this a ``legitimate interest'' that data controllers should safeguard. This interpretation is favored by paragraph 71 of the GDPR's recitals, which suggests the legislative intent was to give the data subject a right "to obtain an explanation of the decision reached" \cite {GDPR-Recital71}. \cite{Goodman2017} argues that GDPR does indeed establish a right to explanation.

On the other hand, it has been argued that, since recitals have no operative effect beyond aiding courts in interpreting ambiguous portions of law, the recital should be ignored because ``[t]here are no ambiguities in the language that would require further interpretation with regard to the minimum requirements that must be met by
data controllers'' \cite{Wachter2017}. This prediction is plausible, but as the authors themselves admit, it is ``only one possible future'' that depends on European courts maintaining the same course of statutory interpretation during a technological and societal sea change. And even in this strict reading of the GDPR, where each article is read in isolation and without reference to recitals, the authors deduce, at minimum, a ``right to explanation of the functionality of the system, a so-called 'right to be informed.'''

\subsubsection{AI Liability Directive}

The final and potentially the most potent source of a right to explainability springs from the AI Liability Directive \cite{AILiability}. In this proposed directive, the European Parliament and Council have laid out a burden-shifting scheme applying to fault-based liability claims for damages resulting from the output of AI systems. To avoid the harshest legal outcomes under this proposed burden-shifting, deployers of AI models of all risk profiles should make their systems transparent and explainable. 

Generally, establishing fault-based liability requires the claimant to prove that the defendant's fault has caused their damages. However, the AI Liability Directive would create under some conditions a presumption that a claimant's damages have been caused by the defect or fault in the AI system deployed by the defendant. This effectively shifts the burden to the defendant to prove that their system \emph{did not} cause the claimed damages. This policy is an acknowledgment that placing the burden of proving causation on the claimant would drastically reduce their chances of recovery in almost all cases because of the inherent difficulty in ``connecting the dots'' between an AI system's inner workings and its outputs. 

The presumption of causation always applies to claims against high-risk AI systems, except ``where the defendant demonstrates that sufficient evidence and expertise is reasonably accessible for the
claimant to prove the causal link'' \cite{AILiability}(art 4.4). To limit the risk of liability, the deployers of high-risk models should ensure that they are not using black box models, and publish documentation and other assets that adequately explain the high-risk model's behavior. Doing so would presumably shift the burden of causation back to potential claimants.

For non-high-risk models, the presumption of causality would only apply where, according to Article 4.5, ``the national court considers it excessively difficult for the claimant to prove the causal
link'' \cite{AILiability}. Since there is no consensus as to what would be a scientifically or legally sufficient threshold of explaining the outputs of a generative AI system, it's unclear if or how anyone could establish with clear and convincing evidence that a fault in an AI system has proximately caused its harmful outputs. As a result, national courts in practice might interpret this section (as enacted locally) in a manner more sympathetic to claimants than intended by the EU legislature. This is a key area where again, deployers of non-high risk models can avoid potential liability by pro-actively taking steps to make their systems explainable. 

\vspace{5mm}

\indent Despite the lack of a clear directive requiring explainability from all categories of AI models deployed in Europe, the penumbra of ``a right to explainability'' can reasonably be inferred from several sections and recitals of the AI Act \cite{AIAct}, the GDPR \cite{GDPR}, and the proposed AI Liability Directive \cite{AILiability}. Proposed and existing laws in the United States emphasize explainability as a core requirement for compliant algorithmic systems. For these reasons, companies seeking to limit their legal and regulatory risk in this uncertain environment will likely focus substantial resources on developing and deploying explainable AI models – even if such models do not actually improve accuracy or safety. 

\section{Polysemy of ``Explainability''}
The term ``explainability'' is used in a wide array of scenarios. Similarly, trustworthiness, safety, and related terms are used, often without clear guidelines. While there have been efforts to taxonomize qualities, formats, data types, and purposes of XAI, those conceptual definitions are difficult to apply and may not clearly delineate technical development in practice. Especially seeing from how legislative texts are written, ``explainability'' has become a catch-all phrase, which is up for broad interpretation. For comprehensive reviews and taxonomical organizations, see \cite{Stepin2021,BarredoArrieta2020,Nauta2023,Saeed2023}. Here we highlight five areas that any explainability regulation should address specifically:

First, there are two approaches to explaining AI; \underline{global and local}. Global explanations attempt to answer how the AI model works, as an overall characterization obtained before a specific sample has been processed, \emph{ex ante}. Furthermore, global explanations look at overall relationships between variables, which are especially critical when dealing with non-linear relationships. Local explanations, on the other hand, refer to how the AI model worked on a particular sample \emph{post hoc}. Often, a non-linear relationship can be characterized linearly in a small window (local) of some feature values. Or, in classifying and predicting based on a specific sample image, the model may have used a small subset of features.  

Under our reading, Article 13 of the GDPR \cite{GDPR} would require deployers to provide global \emph{ex ante} explanations at the moment the subject's personal data is collected for the purpose of informing them about the broad characteristics of an automated decision process. Articles 15 and 22 likely reference the need for local \emph{post hoc} explanations upon the request of the subject. Similarly, the Equal Credit Opportunity Act, which requires that adverse automated credit decisions be accompanied by the reasons for the decision, is an example of a requirement for local, \emph{post hoc} explainability.

Third, generally speaking, the current XAI methodology can’t give \underline{comprehensive yet human-understandable} explanations for black box AI. Only the full model and weights can truly describe the ``complete mechanisms'' of deep neural networks. In general, when explanations are complete, they could simply be used as an efficient and interpretable model with equivalent performance. In practice, post-hoc explainability methods substantially reduce the model mechanisms and characterize different aspects of the model, e.g., by estimating and visualizing the relative importance of input features (also known as saliency maps).

Fourth, the necessity and inevitability of simplification when explaining AI becomes a question about \underline{for whom explainability is intended}. There would exist different levels of explanations similar to how complex phenomena may be explained to an expert, educated adult, and high schooler. The target audience and context for explainability should be considered the central element. However, that is currently not the case. When citizens have a right to explanation from high-risk AI systems, should the AI operators provide uniform and identical explanations? Furthermore, the delivery and interactivity of explanations are crucial if they are to be of practical use to the average person. Currently, operationalizing this aspect of XAI is missing.

Practically, the explainability requirements of consumer protection regulations such as the U.S. Equal Credit Opportunity Act \cite{EqualCredit15/1691} and the GDPR \cite{GDPR} imply that the provided explanation should be understandable by the average person. On the other hand, explanations published by corporations hoping to avoid the presumption of causality promulgated under the EU AI Liability Directive \cite{AILiability} would need to be much more detailed. Overcoming a legally mandated presumption relating to causality cannot be achieved through pedestrian means, likely requiring the publication of data sets and methodologies that illuminate exactly how the claimant's damages were (not) caused by the AI system's alleged fault.

Fifth, being able to explain the model's decision-making process is related to the ability to \underline{control and interact with the AI model} in a meaningful way \cite{Teso2019}\cite{Schmid2020}. While explanations may (truthfully) characterize the model, how to change its behavior according to our desire is not always feasible. One may correct a misbehavior (e.g., misclassification) by re-labeling data, re-training the model, and even including certain pre- and post-processing steps, but this remains an unsolved challenge. We may encourage the model to work closer to how we want it to behave, but short of compiling hard-coded rules\footnote{Rule-based algorithms exist and are useful in a wide range of areas. Those algorithms are not considered AI in this position paper.}, there's no guarantee these guardrails will be effective.  

The aforementioned five aspects of explainability are closely related. For example, local explanations may be reasonable and faithful for a given sample; however, they may not provide sufficient information to control the model. Furthermore, the simplification of AI's decision-making processes, which is necessary for most complex non-linear models (e.g., DNNs), must be informed by the target audience. Beyond these five categories, there exist other definitions of explainability. Mechanistic explanations focus on parts of the model (e.g., layers, sub-networks), whereas functional explanations are about the model's functions \cite{Paez2019}. \cite{Ginsberg1986,Grahne1998} examine counterfactual explanations which can be used to explain black box models. Overall, these different interpretations of explainability would significantly impact what must be explained, assuming the technical challenges are solved. Even with the best intentions and improved technical capabilities, misalignment between the intention of laws and the implementation of AI companies will create confusing and uninformative processes. 

\section{Failure Modes of XAI}
Obtaining explanations from black box AI models remains highly challenging, with a number of competing methods and algorithms. There are multiple ways that explainers can give misleading information. Here, we present five critical failure modes for XAI, that are applicable to DNNs and beyond.

\begin{figure}[tb!]
\centering
\includegraphics[width=.5\textwidth]{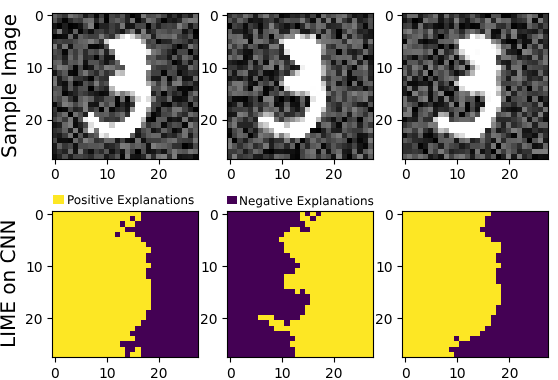}
  \caption{The first row shows the same image of 3, with different noise. CNN trained on MNIST correctly classifies these sample images as `3'. Explanations from LIME are shown in the second row.}\label{fig:mnist}
\end{figure}

\subsubsection{Robustness}
First, XAI methods may not be robust to small changes in inputs \cite{AlvarezMelis2018}. For example, in \autoref{fig:mnist}, three samples are based on the same image of `3'. For this experiment, small and independent noise is added. A simple neural network with one convolutional layer is trained on the MNIST database \cite{Deng2012}. The trained CNN model correctly classifies all three images with high confidence. Nonetheless, explanations by Local Interpretable Model-agnostic Explanations (LIME; on the second row) \cite{LIME} are substantially varied. The positive and negative influence of pixels, as estimated \autoref{fig:mnist}, would present drastically different interpretations of how the classification has been made.

Relatedly, \cite{Adebayo2018} investigated the robustness of saliency maps, which are designed to quantify and highlight important pixels with respect to the class label. They demonstrated that many advanced saliency maps show similar explanations even with data randomization. In the worst cases, some saliency maps, which are supposed to provide local explanations based on the model, resemble edge detection, which is by design independent of the model or the classification \cite{Adebayo2018}. Generally, these types of problems may be mitigated by optimizing for robustness \cite{AlvarezMelis2018} or selecting appropriate methods \cite{Adebayo2018}. 

\subsubsection{Adversarial Attacks}
Second, adversarial attacks on XAI have been investigated, such that even well-performing methods give nonsensical explanations \cite{Dombrowski2019}. Instead of independent noise added in the first failure mode, the attackers can specifically design imperceptible changes to the input (e.g., the image looks identical to a human before and after modification), which completely change the explanations. This type of vulnerability mirrors well-known adversarial attacks on DNNs where carefully designed imperceptible changes to the input can cause the model to misbehave and misclassify. Mitigation strategies for adversarial attacks include explanation aggregation and model regularization \cite{Baniecki2023}. Nonetheless, attacks and defenses surrounding explanations will continue to evolve as XAI becomes more mainstream and public-facing.

\begin{figure}[tb!]
\centering
\includegraphics[width=.5\textwidth]{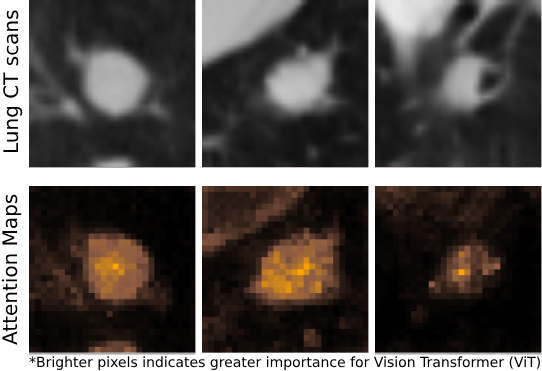}
  \caption{ViT model is fine-tuned on lung CT scans, to classify benign vs. malignant tumors. Attention maps are popular explanations for ViT.}\label{fig:lung}
\end{figure}

\subsubsection{Partial Explanations}
Third, explanations can be \emph{too incomplete} to give real insight. Explanations of complex AI models such as DNNs necessitate simplification of underlying mechanisms. While there are multiple evaluation frameworks to measure the faithfulness of XAI methods \cite{Brocki2023fidelity,Brocki2023,Dawoud2023}, these measures are nonetheless relative without ground truth. Explaining a small fraction of complex behaviors may lead to over-confidence in both AI and XAI models. For example, we fine-tuned the ViT model from pre-trained DINO \cite{Caron2021}, on lung CT scans \cite{Armato2011}. The model performs well classifying benign vs. malignant cases with an accuracy $86\%$. When we obtained attention maps (i.e., de facto standard explanations for ViT), we observe that they highlight lung nodules (\autoref{fig:lung}). At first glance, clinicians may be content with these explanations as AI has seemingly used the tumors for classification. However, without further distinction (e.g., higher resolution; signed importance), we don’t actually learn which parts of a tumor are used for the model's inference. Saliency maps can tell us where the model is looking but not what the model sees \cite{Rudin2019}. 

\begin{figure*}[tb]
\centering
\includegraphics[width=\textwidth]{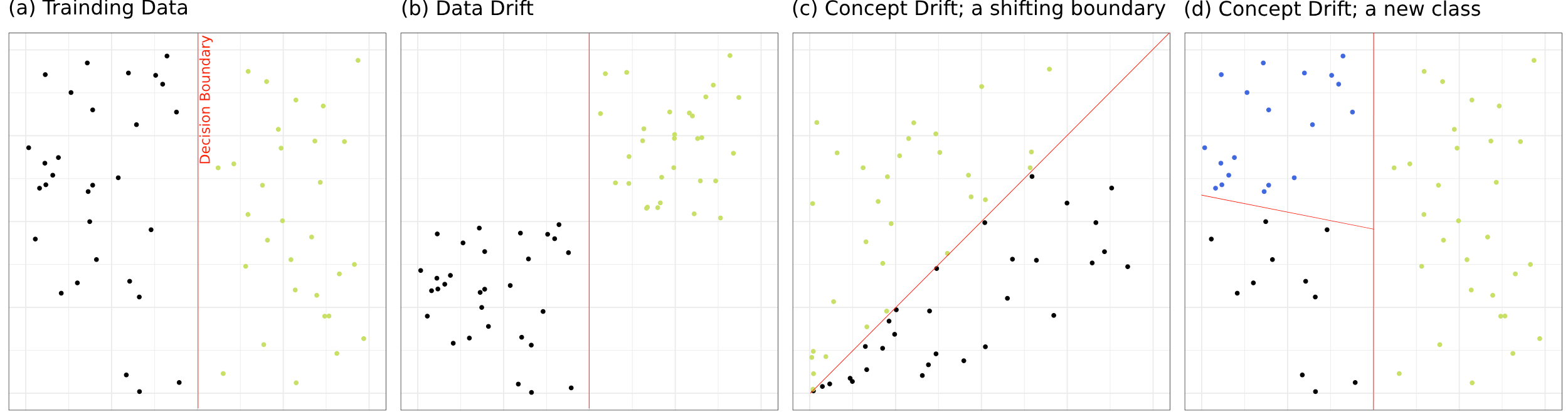}
  \caption{\emph{(a)} The training data is initially used to produce a model, whose decision boundary classifies data in production. \emph{(b)} Data drift occurs when the distribution of data has significantly changed. \emph{(c)} The relationship between data and concept (i.e., class) may change over time, resulting in a concept drift. \emph{(d)} Concept drift could arise from introduction of a new class, that was not included in the training data.}\label{fig:drift}
\end{figure*}

\subsubsection{Data and Concept Drift}
Fourth, data drift \cite{Maimon2010} and concept drift \cite{Widmer1996} refer to the change in data distribution and in applicable domains, respectively. Once trained and deployed, AI systems with explainers tend to remain fixed. Nonetheless, data and their relationship to underlying concepts fed to the systems may change over time (\autoref{fig:drift}). For example, an AI model trained with data from African hospitals may end up being used in Asian countries. Patients of a different ethnicity impact how AI is performing, as well as what explanations are given. XAI that worked faithfully in Africa might gives misleading explanations in Asia. On the other hand, over time, new concepts may appear, that originally did not exist in the training data (\autoref{fig:drift}(d)). For example, malignant tumors among Asian patients might exhibit novel characteristics in medical images, unseen among African patients. Then, XAI methods that were trained and evaluated on African populations might not work on medical images from Asian patients. To overcome these challenges, we must evaluate XAI for the application under consideration and continuously update our methods over time. Continuous data monitoring, collection, training, evaluation and deployment would involve significant costs and effort.

\subsubsection{Anthropomorphization}
Fifth, humans tend to anthropomorphize explanations by assuming a closer alignment of the model’s and human’s inference process than is actually the case. Starting with the birth of neural networks, an implicit or explicit connection between ``artificial'' and ``biological'' is prevalent. Mistaking that DNNs work like our brains wrongly gives an impression that they must be utilizing similar features and logic to arrive at their final decision \cite{Inie2024}. For example, when applied on natural images where objects are classified, humans may like explanations that focuses on regions of classified objects. However, it is entirely possible that the AI classifier, in fact, derives information from background pixels. Nonetheless, this type of erroneous qualitative evaluation occurs regularly, as ground truth on explanations is fundamentally lacking. Anthropomorphization is particularly problematic for XAI as the seemingly plausible explanations can easily trick humans to believe in the system.
 
Overall, these failure modes demonstrate the great risks that arise from blindly adapting XAI methods. At the very minimum, careful consideration and evaluation for applications are needed to reduce erroneous conclusions from explanations. There are likely other potential pitfalls, while more robust and complete XAI methods are being developed. Nonetheless, in our perspective, these failure modes will continue to persist in the current paradigm of XAI.

\section{Market Conditions}
In the market, both established and start-up companies are rushing to offer XAI solutions. Most of the products that we have surveyed do not interact with the underlying AI model in a meaningful way. The concept of XAI is already co-opted to provide false sense of security, where turn-key products added on a top of existing AI infrastructure may seem sufficient to end-users.


Arthur AI, Inc. claims to be ``the very first company to launch ML performance and data drift monitoring for computer vision models, including in-depth explainability'' \cite{ArthurWhitePaper}. As one of the most well-funded start up in this space, its Series B funding was US$\$42$M in 2022. As a specific example, for its computer vision product, it markets ``Explainability and Monitoring for Your CV Models''. Their documentation repeatedly states ``explainability'' without providing details, let alone advantages and disadvantages\footnote{\url{https://docs.arthur.ai/docs} Accessed April 26, 2024}. In the Arthur SDK (Software Development Kit) Python API (Application Programming Interface) Reference\footnote{\url{https://sdk.docs.arthur.ai/} Accessed April 26, 2024}, it becomes clear that Arthur implemented LIME (Local Interpretable Model-Agnostic Explanations) under the hood. While LIME is one of the most popular XAI methods, its limitations should be up front to ensure appropriate applications.

Google is one of the largest and the most advanced companies in AI with a market capitalization of $\sim$US$\$2$ trillion. Google would advertise Vertex AI\footnote{\url{https://cloud.google.com/vertex-ai/docs/explainable-ai/overview} Accessed April 26, 2024} with a number of explainability methods, which claim to be authoritative and to avoid the pitfalls of black box AI. Information on the limitations of their methods seems to focus on narrow aspects, instead of the fundamental intractable issue of attempting to explain a black box. While Google DeepMind publishes excellent methods and papers on XAI that dive deeper into technical details, potential clients likely do not read academic publications. 

Note that over-confident marketing of XAI is not at all unique to these two companies. Explainability methods have been developed in a research environment, often within narrowly defined applications and assumptions. As those methods are packaged in a for-profit environment, technical nuances may have been glanced over.

\section{Discussion and Conclusion}
Considering the variety of XAI methods with a wide range of goals and operating characteristics, the use of these perfunctory explanation methods and flawed compliance tools may trigger many failure modes of XAI and provide misleading or unreliable explanations. This affects the following three groups interfacing with AI systems. Feedback loops among these groups amplify misleading impressions about XAI:

\underline{Regulators and politicians} are under the impression that explainability is an issue of implementation, instead of a fundamental research problem. Law and policy documents fail to specify what are and how to obtain explanations, while simultaneously requiring certain AI operators to provide explanations. The AI Act in EU \cite{AIAct} and the Executive Order in the USA \cite{EO} demand safe, trustworthy, and explainable AI. These will force tech companies to implement and outsource explainability for their AI systems, regardless of their accuracy, reliability, or utility.

\underline{AI developers and operators} may view “explanations” given by existing XAI methods as golden standards. Under legislative and economic pressures, companies utilizing AI systems may adapt suboptimal XAI that are not meaningfully explaining the decision making process. Explanations that seem to match human expectations may garner approval, and companies will become falsely confident that their systems are working as intended. Without detailed regulations that define and evaluate standards for XAI, ad hoc implementations could lead a right to explanation astray.

\underline{Clients and citizens} are conditioned to rely and trust the AI systems on the basis of sweeping laws and AI systems promising explanations.  Whereas a black box AI raises an inherent concern about their unknowable operating characteristics, legally mandated and confidently marketed XAI, is provided with a false aura of trustworthiness. User studies show inclusion of explanations increases user trust and decreases decision-making time. But at the same time, XAI (specifically, LIME \cite{LIME}, SHAP \cite{Lundberg2017}, and TreeInterpreter \cite{Saabas2014}) substantially decreased the decision accuracy of human analysts in a fraud detection study \cite{Jesus2021}. Regardless of the model performance, users of AI become more reliant when explanations are provided \cite{BritoDuarte2023}. 

From both technical and legislative perspectives, explainability plays an important role in AI governance. While it is intuitive to present and advocate for the need to understand, interpret, and explain a model's decision-making process, the methodologies of XAI require thorough consideration of their applications, goals, and failure modes. We must dispel the false sense of security in XAI arising from over-confident products that seemingly satisfy legislative and policy texts that vaguely imply a right to explanation. 

\appendix

\section*{Ethical Statement}

There are no ethical issues.

\section*{Acknowledgments}


\bibliographystyle{named}
\bibliography{XAI-FalseSense}

\end{document}